\documentclass[journal,twocolumn,10pt]{IEEEtran}
\usepackage{graphicx}
\usepackage{amssymb}
\usepackage{amsmath}
\usepackage{mathtools}
\usepackage{dsfont}
\usepackage{cite}
\usepackage{stfloats}
\usepackage{subfigure}
\usepackage{psfrag}
\usepackage[mathscr]{euscript}
\usepackage{acronym}  
\usepackage{booktabs}
\usepackage{bbm}
\usepackage[numbers,sort&compress]{natbib}
\usepackage{algorithmic}
\usepackage{algorithm}

\usepackage{float}
\usepackage{balance}

\acrodef{CCDF}{complementary cumulative distribution function}
\acrodef{CF}{characteristic function}
\acrodef{PPP}{Poisson point processe}
\acrodef{RV}{random variable}
\acrodef{i.i.d.}{independent and identically distributed}
\acrodef{PDF}{probability distribution function}
\acrodef{CDF}{cumulative distribution function}
\acrodef{ch.f.}{characteristic function}
\acrodef{AWGN}{additive white Gaussian noise}
\acrodef{SNR}{signal-to-noise ratio}
\acrodef{LRT}{likelihood ratio test}
\acrodef{DRT}{distance ratio test}
\acrodef{GLRT}{generalized likelihood ratio test}
\acrodef{CRLB}{Cram\'{e}r-Rao lower bound}
\acrodef{CRB}{Cram\'{e}r-Rao bound}
\acrodef{ZZLB}{Ziv-Zakai lower bound}
\acrodef{ZZB}{Ziv-Zakai bound}
\acrodef{LOS}{line-of-sight}
\acrodef{ToF}{time-of-flight}
\acrodef{NLOS}{non-line-of-sight}
\acrodef{GDOP}{geometric dilution of precision}
\acrodef{GPS}{Global Positioning System}
\acrodef{FIM}{Fisher information matrix}
\acrodef{PEB}{position error bound}
\acrodef{SPEB}{squared position error bound}
\acrodef{TOA}{time-of-arrival}
\acrodef{TOF}{time-of-flight}
\acrodef{WSN}{wireless sensor network}
\acrodef{MAC}{medium access control}
\acrodef{RSS}{received signal strength}
\acrodef{WAF}{wall attenuation factor}
\acrodef{TDOA}{time difference-of-arrival}
\acrodef{RF}{radiofrequency}
\acrodef{RTT}{round-trip time}
\acrodef{AOA}{angle-of-arrival}
\acrodef{MF}{matched filter}
\acrodef{ED}{energy detector}
\acrodef{ML}{maximum likelihood}
\acrodef{MSE}{mean-square error}
\acrodef{RMSE}{root-mean-square error}
\acrodef{LEO}{localization error outage}
\acrodef{ppm}{part-per-million}
\acrodef{ACK}{acknowledge}
\acrodef{UWB}{Ultrawide bandwidth}
\acrodef{TNR}{threshold-to-noise ratio}
\acrodef{LS}{least squares}
\acrodef{IR-UWB}{impulse radio UWB}
\acrodef{FCC}{Federal Communications Commission}
\acrodef{TH}{time-hopping}
\acrodef{PPM}{pulse position modulation}
\acrodef{MUI}{multi-user interference}
\acrodef{PDP}{power delay profile}
\acrodef{BPZF}{band-pass zonal filter}
\acrodef{SIR}{signal-to-interference ratio}
\acrodef{SINR}{signal-to-interference-plus-noise ratio}
\acrodef{RFID}{radio frequency identification}
\acrodef{WPAN}{wireless personal area network}
\acrodef{WWB}{Weiss-Weinstein bound}
\acrodef{DP}{direct path}
\acrodef{MF}{matched filter}
\acrodef{MMSE}{minimum-mean-square-error}
\acrodef{SBS}{serial backward search}
\acrodef{SBSMC}{serial backward search for multiple clusters}
\acrodef{NBI}{narrowband interference}
\acrodef{WBI}{wideband interference}
\acrodef{INR}{interference-to-noise ratio}
\acrodef{CR}{channel response}
\acrodef{CIR}{channel impulse response}
\acrodef{CR}{channel  response}
\acrodef{RADAR}{radar}
\acrodef{MUR}{Multistatic radar}
\acrodef{JBSF}{jump back and search forward}
\acrodef{HDSA}{high-definition situation-aware}
\acrodef{RRC}{root raised cosine}
\acrodef{ST}{simple thresholding}
\acrodef{BTB}{Bellini-Tartara bound}
\acrodef{P-Max}{$P$-Max}  
\acrodef{MIMO}{multiple-input multiple-output}
\acrodef{MAP}{maximum a posteriori}
\acrodef{FG}{factor graph}
\acrodef{OP}{outage probability}
\acrodef{WED}{wall extra delay}
\acrodef{RMS}{root mean square}
\acrodef{SPAWN}{sum-product algorithm over a wireless network}
\acrodef{MDD}{minimum distance distribution}
\acrodef{MAP}{maximum a posteriori probability}
\acrodef{SAP}{small cell access point}
\acrodef{UE}{user equipment}
\acrodef{MBS}{macro cell base station}
\acrodef{UER}{\ac{UE} Relay}
\acrodef{D2D}{device-to-device}
\acrodef{MBS}{macro base station}
\acrodef{CSI}{channel state information}
\acrodef{OGR}{outage guard region}
\acrodef{FUR}{feasible UER region}
\acrodef{EHR}{energy harvesting region}
\acrodef{EH}{energy harvesting}
\acrodef{D2D-EHSN}{D2D communication provided \ac{EH} small cell network}
\acrodef{D2D-EHHN}{D2D communication provided \ac{EH} heterogeneous network}
\acrodef{3GPP}{3rd Generation Partnership Project}
\acrodef{BS}{base station}
\acrodef{DF}{decode and forward}
\acrodef{CCDF}{complementary cumulative distribution function}
\acrodef{ZF}{zero forcing}
\acrodef{RZF}{regularized zero forcing}
\acrodef{WLLN}{weak law of large number}
\acrodef{SLLN}{strong law of large numbers}
\acrodef{TDD}{Time-division duplex}
\acrodef{EE}{energy efficiency} 
\acrodef{HetNet}{heterogeneous network} 
\acrodef{SCP}{Single Cell Processing}
\acrodef{CBF}{Coordinated Beamforming}
\usepackage{color}
\usepackage{dsfont}
\usepackage{bbm}

\def\PST{P_{\mathrm{st}}}
\def\PUT{P_{\mathrm{ut}}}

\def\LS{\lambda_{\mathrm{s}}}
\def\LU{\lambda_{\mathrm{u}}}

\DeclareMathAlphabet{\mathsf}{OML}{cmbr}{m}{it}

\newtheorem{definition}{\bf Definition}

\newcommand{\bd}{\begin{description}}
\newcommand{\ed}{\end{description}}
\newcommand{\be}{\begin{enumerate}}
\newcommand{\ee}{\end{enumerate}}
\newcommand{\bi}{\begin{itemize}}
\newcommand{\ei}{\end{itemize}}
\newcommand{\bl}{\begin{list}}
\newcommand{\el}{\end{list}}
\newcommand{\bt}{\begin{tabbing}}
\newcommand{\et}{\end{tabbing}}

\newcommand{\paperTitle}{Boosting Dynamic TDD in Small Cell Networks by the Multiplicative Weight Update Method}

\begin{document}

{
\title{\paperTitle}

\author{

	\vspace{0.2cm}
	    Jiaqi~Zhu$^\dagger$, 
            Nikolaos Pappas$^\mathsection$,
            and 
            Howard~H.~Yang$^\dagger$ \\
       
		$^\dagger$ \textit{ZJU-UIUC Institute, Zhejiang University, China }\\
          $^\mathsection$ \textit{Department of Computer and Information Science, Linköping University, Sweden}
          
\thanks{The work of J. Zhu and H.~H.~Yang was supported in part by the National Natural Science Foundation of China under Grant 62201504, in part by the Zhejiang Provincial Natural Science Foundation of China under Grant LGJ22F010001, and in part by the Zhejiang – Singapore Innovation and AI Joint Research Lab. The work of N. Pappas has been supported by the Swedish Research Council (VR), ELLIIT, the European Union (ETHER, 101096526), and the European Union's Horizon Europe research and innovation programme under the Marie Skłodowska-Curie Grant Agreement No. 101131481 (SOVEREIGN).}  

}
\maketitle
\acresetall
\thispagestyle{empty}
\begin{abstract}
We leverage the Multiplicative Weight Update (MWU) method to develop a decentralized algorithm that significantly improves the performance of dynamic time division duplexing (D-TDD) in small cell networks.
The proposed algorithm adaptively adjusts the time portion allocated to uplink (UL) and downlink (DL) transmissions at every node during each scheduled time slot, aligning the packet transmissions toward the most appropriate link directions according to the feedback of signal-to-interference ratio information. 
Our simulation results reveal that compared to the (conventional) fixed configuration of UL/DL transmission probabilities in D-TDD, incorporating MWU into D-TDD brings about a two-fold improvement of mean packet throughput in the DL and a three-fold improvement of the same performance metric in the UL, resulting in the D-TDD even outperforming Static-TDD in the UL.
It also shows that the proposed scheme maintains a consistent performance gain in the presence of an ascending traffic load, validating its effectiveness in boosting the network performance. 
This work also demonstrates an approach that accounts for algorithmic considerations at the forefront when solving stochastic problems.
\end{abstract}

\acresetall

\section{Introduction}\label{sec:intro} 
Time-division duplex (TDD) is a communication protocol under which the transmissions of network nodes take place over non-overlapping time slots but in the same frequency band. 
TDD has been widely used in 3G, 4G, and 5G systems, enabling low complexity and scalable control over the flow of uplink (UL) and downlink (DL) traffic at the nodes. 

Typically, TDD systems accommodate the UL/DL traffic asymmetry by adjusting the portion of time slots allocated to UL and DL transmission \cite{HolHarTos:11, 3GPP:10, QoMEX2015, XenakisWCNC2016}.
Depending on the configuration of UL/DL transmissions, TDD schemes can be broadly categorized into two classes, i.e., static TDD (S-TDD) and dynamic TDD (D-TDD). 
S-TDD requires synchronization of all UL and DL node activities, where during every communication round, the transmissions across all the nodes in the network are aligned in either UL or DL directions. 
In contrast, D-TDD allows each node to configure its subframe to accommodate whichever link direction needs it the most \cite{SheKhoEri:12}. Consequently, D-TDD holds the potential for higher spectrum utilization and reduced latency, making it particularly appealing for network scenarios with significant traffic fluctuation. System-level comparisons between S-TDD and D-TDD have been conducted, assessing factors such as coverage probability \cite{ZhoCheWan:15}, achievable rate \cite{GupKul:16}, and energy efficiency \cite{SunSheWil:16}, showing the gains attainable by D-TDD. These comparisons consistently demonstrate the advantages offered by D-TDD. As a result, D-TDD schemes have attracted considerable research interest \cite{DinDinMao:18, DinLopXue:16d, LiHuaSha:18, YuYanIsh:15, RazZlaPak:20}. 

Despite the salient advantages, D-TDD suffers additional inter-cell interference introduced by asynchronous UL/DL transmissions, hindering its implementation in densely deployed small cell networks \cite{LopDinCla:15}. 
As pointed out by \cite{YanGerZho:17}, S-TDD outperforms D-TDD in the UL, while the reverse holds for DL operations. 
This phenomenon chiefly ascribes to the asymmetric transmissions in D-TDD, which reduce DL interference at the expense of increasing UL interference. 
Furthermore, \cite{YanGerZho:17} discloses that the number of scheduled \acp{UE} can significantly affect network performance.

The Multiplicative Weight Update (MWU) algorithm is a well-known meta-algorithm that uses the multiplicative update rule to change weights iteratively. The simplest use case is the problem of prediction from expert advice, in which a decision maker iteratively selects an expert’s advice to follow. 
The method assigns initial weights to the experts (usually identical initial weights) and updates these weights iteratively according to the feedback of how well an expert performed: reducing it in case of poor performance and increasing it otherwise \cite{AroHazKal:12}. 
Albeit primarily employed to tackle constrained optimization problems, the multiplicative weights method appears well-suited to address the issue of allocating the time slots to asynchronous UL and DL transmission in the D-TDD scheme.

In this paper, we showcase how to leverage the MWU algorithm to enhance the D-TDD scheme by adaptively updating the time portion allocated to DL and UL over time. 
We evaluate the performance of small cell deployments under the proposed D-TDD with MWU, comparing with D-TDD and S-TDD. Our numerical results show that the use of MWU in D-TDD brings about a two-fold improvement in the DL and a three-fold improvement in the UL, which achieves comparable or even better performance of S-TDD in the UL. We also find the outstanding performance of the proposed D-TDD scheme when subject to the maximum number of served UEs per SAP. Also, this work tries to keep algorithmic considerations at the forefront when solving stochastic problems.


\section{System Model}\label{sec:sysmod}
This section details the network topology, traffic model, and radio access schemes. 

\subsection{Network Topology}

We consider a small cell network consisting of \acp{SAP} and UEs, whose spatial locations follow independent \acp{PPP} $\Phi_{\mathrm{s}}$  and $\Phi_{\mathrm{u}}$, with spatial densities  $\LS$ and $\LU$, respectively.
All SAPs and UEs are equipped with a single antenna and transmit with power $\PST$ and $\PUT$, respectively.{\footnote{Note that PPP is a widely used stochastic model to characterize the spatial deployment of cellular networks.}}
We assume the channels between any pair of nodes to be narrowband and affected by two attenuation factors: small-scale Rayleigh fading and large-scale path loss. 
In this network, the UEs associate with the nearest SAPs in space, corresponding to the highest average received power (in the downlink) association policy. 
Such an association can result in multiple UEs connecting to one \ac{SAP}, we limit the maximum number of UEs served by each \ac{SAP} (denoted by $N_{\mathrm{s}}$) to $K_{\mathrm{s}}$, and assume that each SAP randomly selects one of the UEs in its coverage to serve at each time slot.

\subsection{Traffic Pattern}
We model the traffic profile by a discrete-time queuing system, where time is segmented into slots with equal duration. 
We assume all temporal dynamics, i.e., packet arrivals and departures, take place at each time slot, and the transmission of a data packet occupies one time slot. 
For a generic UE, we model its UL/DL packet arrivals as independent Bernoulli processes with rates $\xi_{\mathrm{U}}, \xi_{\mathrm{D}} \in [0,1]$ (packet/slot) \cite{YanGerQue:16}. We further assume that each UE accumulates all incoming packets in an infinite-size buffer. At the same time, every SAP maintains $K_\mathrm{s}$ distinct buffers to store the packets for each intended UE in the downlink. 

At the beginning of each time slot, a scheduled transmitter (which may be a UE or SAP) sends out the data packet, if any, from the head of its queue. 
At the end of the same time slot, if the SIR at the receiver surpasses a decoding threshold, denoted by $\theta$, the packet is successfully received, upon which the receiver feeds back an \ac{ACK}, and the corresponding packet can be removed from the transmitter's buffer; 
otherwise, the receiver sends a negative acknowledgment (NACK), and the packet will be retransmitted at the next available time slot. 
Consequently, correlations will be amongst the transmitters' buffer states, commonly known as the \textit{spatially interacting queues} \cite{SanBac:17}.


\subsection{Channel Access Scheme}
\begin{figure}[t!]
  \centering{}

    {\includegraphics[width=0.85\columnwidth]{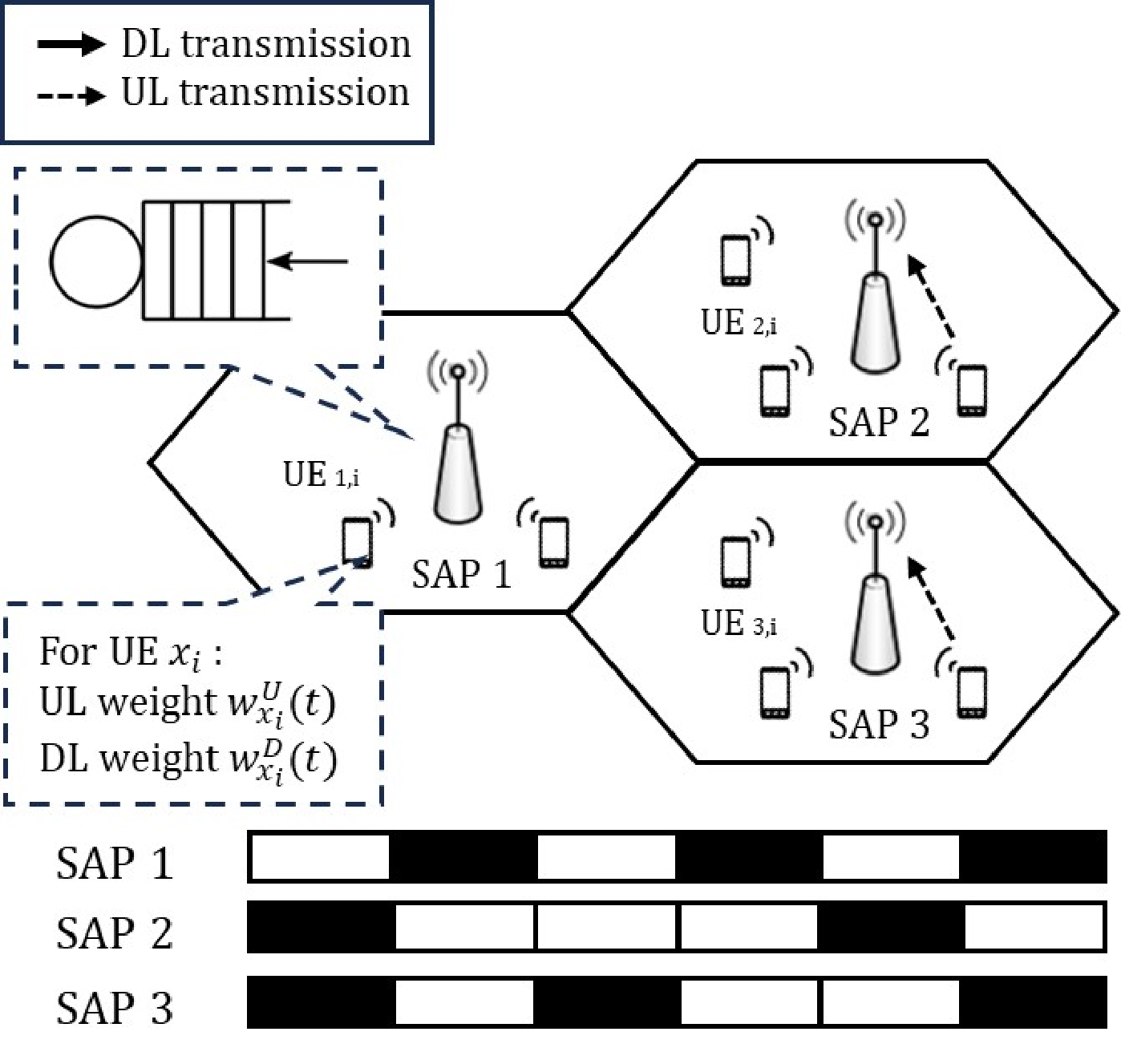}}

  \caption{Illustration of a small cell network with random traffic arrivals. D-TDD with Multiplicative Weight Update (MWU) algorithm is adopted in this example, where SAPs schedule their uplink (white block) and downlink (black block) transmissions independently at each time slot.}
  \label{fig:SysMod_M1}
\end{figure}

To cope with the (asymmetric) traffic dynamic in uplink and downlink, we employ TDD-based policies for radio channel access.
Specifically, we consider two TDD modes of operation for radio access, i.e., S-TDD and D-TDD, described as follows \cite{SheKhoEri:12}.

\subsubsection{S-TDD} At each time slot, all \acp{SAP} transmit either in DL or in UL with probabilities $\eta_\mathrm{S}$ and $1-\eta_\mathrm{S}$, respectively.
\subsubsection{D-TDD} \acp{SAP} independently schedule their transmissions. In a given time slot, a typical \ac{SAP} located at $x_i \in \Phi_{\mathrm{s}}$ transmits in DL (resp. UL) with probability $\eta_{x_i}$ (resp. $1- \eta_{x_i}$).

In the conventional setup, most of the work emphasizes D-TDD to use a fixed configuration of UL-DL proportion, i.e., stipulating $\eta_{x_i} = \eta_\mathrm{D}$, $x_i \in \Phi_{\mathrm{s}}$.
In the following section, we will introduce an algorithm that adaptively updates the time portion allocated to uplink and downlink based on the dynamics of communication environments. 

\subsection{Performance Metric}

In this paper, we employ the performance metric of the \textit{(network) mean packet throughput}, i.e., the number of successfully transmitted packets per time slot. 
A formal definition is given as follows. 

\begin{definition}
\textit{
Let $A_x(t)$ be the number of packets arrived at a typical transmitter $x$ within period $[0,t]$, and $D_{i, x}$ be the number of time slots between the arrival of the $i$-th packet and its successful delivery. The mean packet throughput is defined as
\begin{align}\label{equ:Gen_PktThrPut}
\mathcal{T} \triangleq \lim_{R \rightarrow \infty} \frac{\sum_{x \in \Phi \cap B(0,R)}  \lim\limits_{t \rightarrow \infty} \frac{A_x(t)}{\sum_{i=1}^{A_x(t)} D_{i, x} }}{\sum_{x \in \Phi} \chi{\{ x \in B( 0, R ) \}} }
\end{align}
where $\chi_{\{ \cdot\}}$ is the indicator function, and $B(0,R)$ is a circle centered at the origin with radius $R$.
}
\end{definition}

Note that $D_{i,x}$ in \eqref{equ:Gen_PktThrPut} represents the number of time slots required to successfully deliver the $i$-th packet, and its value is affected by: $(i)$ queueing delay, caused by other accumulated unsent packets, and $(ii)$ transmission delay, due to link failure and retransmission.
By averaging over all nodes, \eqref{equ:Gen_PktThrPut} provides information on the packet throughput across the network.

\section{Multiplicative Weight Update Algorithm for TDD Configuration}
This section presents a decentralized algorithm that leverages the MWU method to adjust the UL/DL transmission probabilities at each node upon every time slot to enhance the network's mean packet throughput. 
Since the design of the algorithm relies on the SIR information\footnote{This paper concentrates on the interference-limited regime where thermal noise is negligible, although the framework can be extended to consider a more general SINR metric. } of each (scheduled) transmission link, we start with a formal characterization of the SIR at each node. 

\subsection{Signal-to-Interference Ratio (SIR)}
Let $\zeta_{x,t} \in \{0,1\}$ be an indicator showing whether a node located at $x \in \Phi \triangleq \Phi_{\mathrm{s}} \cup \Phi_{\mathrm{u}}$ is transmitting at time slot $t$ ($\zeta_{x,t} = 1$) or not ($\zeta_{x,t}=0$). 
By Slivnyark's theorem, we can focus on a typical UE located at the origin with its tagged SAP at $x_0$, which we refer to as the typical SAP. 
If the typical SAP transmits a data packet to the UE at time slot $t$, the received DL {SIR} under S-TDD and D-TDD can be respectively written as:
\begin{align}
\gamma_{\mathrm{S}, t}^{\mathrm{D}} &= \frac{ \frac{ \PST h_{x_0} } {\Vert x_0 \Vert^{\alpha} } }{  \sum\limits_{x \in \Phi_{\mathrm{s}}\setminus x_0  } \!\!\!\! \frac{\PST \zeta_{x,t} h_{x}}{ \Vert x \Vert^{\alpha}}   }, \label{eqn:gammaSD}\\
\gamma_{\mathrm{D}, t}^{\mathrm{D}} &= \frac{ \frac{ \PST h_{x_0} } { \Vert x_0 \Vert^{ \alpha} } }{  \sum\limits_{x \in \Phi_{\mathrm{s}}\setminus x_0  } \!\!\!\! \frac{\PST \zeta_{x,t} h_{x}}{ \Vert x \Vert^{\alpha}}  +\!\! \sum\limits_{z \in \Phi_{\mathrm{u}} } \!\!\!\! \frac{ \PUT \zeta_{z,t} h_{z}}{ \Vert z \Vert^{\alpha}} } \label{eqn:gammaDD}
\end{align}
where $h_x$ denotes the small scale fading from node $x$ to the origin, $\Vert \cdot \Vert$ represents the Euclidean norm, and $\alpha$ is the path loss exponent.

Similarly, the UL {SIR} under S-TDD and D-TDD received by a typical \ac{SAP} from UE $z_0$ can be respectively expressed as:
\begin{align}
\gamma_{\mathrm{S}, t}^{\mathrm{U}} &= \frac{ \frac{ \PUT h_{z_0} } { \Vert z_0 \Vert^{\alpha} } }{ \sum_{z \in \Phi_{\mathrm{u}}\setminus z_0  } \frac{\PST \zeta_{z,t} h_{z}}{ \Vert z \Vert^{\alpha}}  },\label{eqn:gammaSU}\\
\gamma_{\mathrm{D}, t}^{\mathrm{U}} &= \frac{ \frac{ \PUT h_{z_0} } {\Vert z_0 \Vert^{ \alpha} } }{ \sum_{z \in \Phi_{\mathrm{u}}\setminus z_0  } \!\!\! \frac{\PST \zeta_{z,t} h_{z}}{ \Vert z \Vert^{\alpha}} + \sum_{x \in \Phi_{\mathrm{s}} } \!\! \frac{\PST \zeta_{x,t} h_{x}}{ \Vert x \Vert^{\alpha}} }.\label{eqn:gammaDU}
\end{align}

\subsection{Multiplicative Weight Update Algorithm}
Availed with the SIR information, we are now ready to present the main algorithm. 
For every UE $x_i \in \Phi_{ \mathrm{u} }$, we introduce $M_{x_i, t}^{\mathrm{U}}$ and $M_{x_i, t}^{\mathrm{D}}$ as penalty functions associated with the UL and/or DL transmission decision at time slot $t$, given by
\begin{align} \label{equ:UL_pnlt}
M_{x_i, t}^{\mathrm{U}} = \theta - \gamma_{x_i, t}^{\mathrm{U}} (1- e^{-\eta Q_{x_i}^{\mathrm{U}}(t)}),\\ \label{equ:DL_pnlt}
M_{x_i, t}^{\mathrm{D}} = \theta - \gamma_{x_i, t}^{\mathrm{D}} (1- e^{-\eta Q_{x_i}^{\mathrm{D}}(t)}),
\end{align}
in which $\theta$ is the decoding threshold, $Q_{x_i}^{\mathrm{U}}(t)$ and $Q_{x_i}^{\mathrm{D}}(t)$ denote the queue length of the UL and DL buffer, respectively, and $\eta$ calibrates the level of degrees the system controlling the buffer length. 
Moreover, we consider the tagged SAP of UE $x_i$ maintains two factors $w_{x_i}^{\mathrm{U}}(t)$ and $w_{x_i}^{\mathrm{D}}(t)$, respectively, indicating the UL and DL weights, where the higher the weight, the more possible transmission is configured in that direction. 
Therefore, the algorithm shall adequately adjust the weights associated with the UL and DL transmission directions according to the dynamics of the communication environment (reflected in the SIR) such that the transmission can be configured toward the (almost) optimal direction.
In that respect, we leverage the MWU algorithm to devise a scheme that dynamically updates the weights $w_{x_i}^{\mathrm{U}}(t)$ and $w_{x_i}^{\mathrm{D}}(t)$ at every UE.
The detailed steps are summarized in Algorithm~1.

Specifically, the weight in UL (as an example) is reduced by a factor of $(1-\delta)$ if the transmission decision incurs a positive penalty, diminishing subsequent transmissions to prefer this direction.
In contrast, the weight is amplified by a factor of $(1+\delta)$ if the decision results in a negative penalty (in other words, the action receives a reward), reinforcing future actions to align with the current one.
The parameter $\delta$ controls the level of aggressiveness on each update.
Moreover, the penalty function \eqref{equ:UL_pnlt} or \eqref{equ:DL_pnlt} is positive if ($a$) the received SIR does not surpass the decoding threshold and/or ($b$) the queue length associated with the transmission direction is too short. 
As such, it adequately balances the link quality and the queue length. 

\begin{figure}[!t]
    \label{alg:MWU}    
    \begin{algorithm}[H]
		\caption{MWU-based D-TDD}
		\begin{algorithmic}[1]		
        \STATE \textbf{initialize} $w_{x_i}^{\mathrm{U}}(0) = w_{x_i}^{\mathrm{D}}(0) = 1$, $\forall x_i \in \Phi_{ \mathrm{u} }$  
        \FOR {time slot $t \in \{0, 1, 2, ... \}$}
            \IF {$x_i$ is selected}
                \STATE Configure the DL transmission probability as
                \begin{align*}
                    \eta_{x_i}(t) = \frac{ w_{x_i}^{\mathrm{D}}(t) }{  w_{x_i}^{\mathrm{U}}(t) + w_{x_i}^{\mathrm{D}}(t)  }        
                \end{align*}
                \IF {transmission is in UL}
                    \STATE Update $w_{x_i}^{\mathrm{U}}(t)$ as
                    \begin{align*}
                    w_{x_i}^{\mathrm{U}}(t \!+\! 1) =\! \begin{cases}
                        w_{x_i}^{\mathrm{U}}(t) (1-\delta)^{M_{x_i, t}^{\mathrm{U}}/\rho} & \text{if } M_{x_i, t}^{\mathrm{U}} \geq 0 \\
                        w_{x_i}^{\mathrm{U}}(t) (1+\delta)^{-M_{x_i, t}^{\mathrm{U}}/\rho} & \text{if } M_{x_i, t}^{\mathrm{U}} < 0
                        \end{cases}
                    \end{align*}
                    where $\delta$ is a control parameter that determines the learning rate of the updating algorithm, and $\rho > 0$ is a hyperparameter
                \ELSIF {transmission is in DL}
                    \STATE Update $w_{x_i}^{\mathrm{D}}(t)$ as
                    \begin{align*}
                    w_{x_i}^{\mathrm{D}}(t \!+\! 1) =\! \begin{cases}
                        w_{x_i}^{\mathrm{D}}(t) (1-\delta)^{M_{x_i, t}^{\mathrm{D}}/\rho} & \text{if } M_{x_i, t}^{\mathrm{D}} \geq0 \\
                        w_{x_i}^{\mathrm{D}}(t) (1+\delta)^{-M_{x_i, t}^{\mathrm{D}}/\rho} & \text{if } M_{x_i, t}^{\mathrm{D}} < 0
                        \end{cases}
                    \end{align*}
                \ENDIF
            \ENDIF
        \ENDFOR        
		\end{algorithmic}
    \end{algorithm}
\end{figure}

\section{Numerical Analysis}


\begin{figure}[t!]
  \centering{}

  \subfigure[DL transmissions]
    {\label{fig:theta-DL}
    \includegraphics[width=0.9\columnwidth]{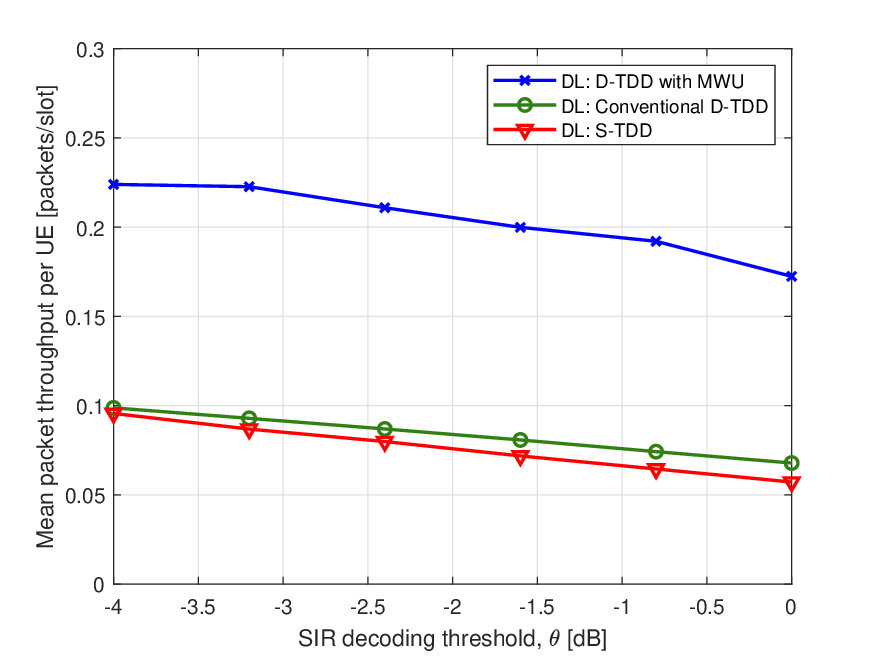}
    }  

  \subfigure[UL transmission]
    {\label{fig:theta-UL}
    \includegraphics[width=0.9\columnwidth]{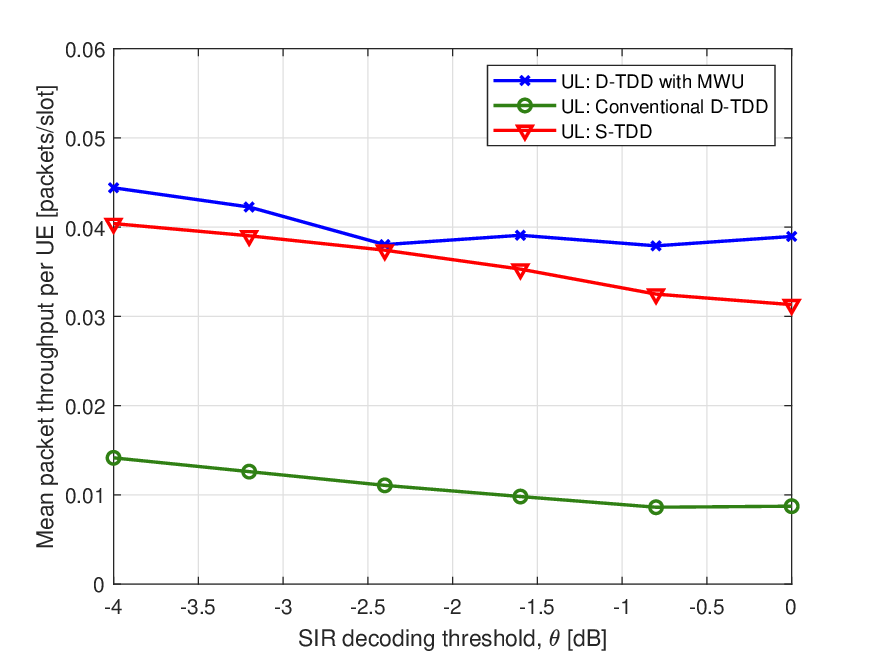}}

  \caption{Mean packet throughput per UE versus SIR threshold: (a) transmissions in downlink and (b) transmissions in uplink.}

  \label{fig:PcktThrpt_vs_theta}
  
\end{figure}

In this section, we conduct simulations to examine the efficacy of the proposed method by comparing the performance of three TDD schemes, i.e., S-TDD, D-TDD under the fixed configuration of UL/DL transmission probabilities, and D-TDD under the MWU method. 
Specifically, we consider a square region with a side length of 1.6 km, where the locations of SAPs and UEs are drawn via independent PPPs with spatial densities $\LS$ and $\LU$, respectively. 
All SAPs and UEs transmit with constant power $\PST$ and $\PUT$, respectively. The maximum number of UEs served by each SAP is limited to $K_{\mathrm{s}}$, and one of the served UEs of each SAP is selected randomly with uniform distribution during each time slot. 
Packets arrive at each node according to the independent Bernoulli process. 
The mean packet throughput is obtained by averaging over $20000$ independent realizations of the above setup. 
Unless otherwise stated, we adopt the following system parameters \cite{YanGerQue:16}: $\LS = 10^{-4}~\mathrm{m}^{-2}$, $\LU = 10^{-3}~\mathrm{m}^{-2}$, $\PST = 2$~ dBm, $\PUT = 17$~dBm, $K_{\mathrm{s}} = 3$, $\theta = 0$~dB, and $\alpha = 3.8$. Moreover, we set the UL packet arrival rate as $\xi_{\mathrm{U}} = 0.05$ and the DL packet arrival rate as $\xi_{\mathrm{D}} = 0.10$.
We assign the same DL time portion for both S-TDD and D-TDD fixed configuration, i.e., $\eta_{x_i} = \eta_\mathrm{S}= \eta_\mathrm{D}$, $\forall x_i \in \Phi_{ \mathrm{s} }$, while D-TDD with MWU updates the value of $\eta_{x_i}$ at each time slot according the algorithm given in Section III-B.

\begin{figure}[t!]
  \centering{}

  \subfigure[DL transmissions]
    {\label{fig:Km-DL}
    \includegraphics[width=0.9\columnwidth]{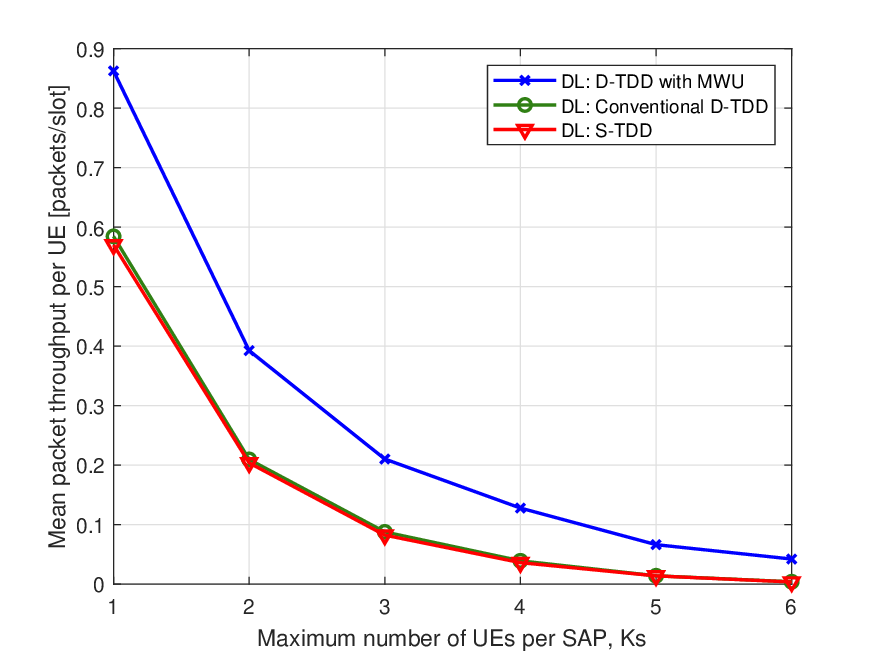}}

  \subfigure[UL transmissions] 
    {\label{fig:Km-UL}
    \includegraphics[width=0.9\columnwidth]{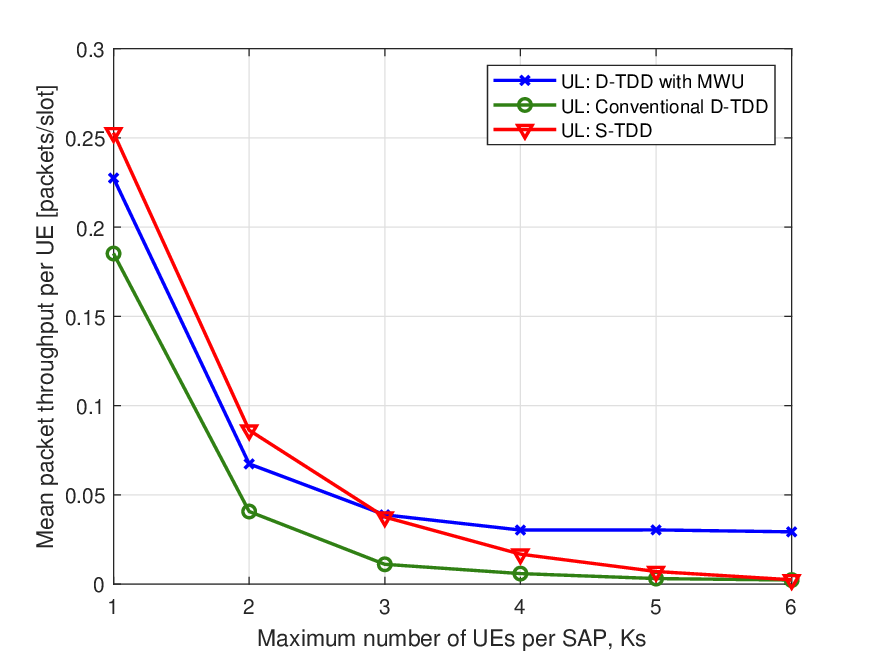}}

  \caption{Mean packet throughput per UE versus number of UEs per SAP: (a) transmissions in downlink and (b) transmissions in uplink.}
  \label{fig:PcktThrpt_vs_UE_per_cell}
  
\end{figure}

\begin{figure}[t!]
  \centering{}

  \subfigure[DL transmissions]
    {\label{fig:delay-DL}
    \includegraphics[width=0.9\columnwidth]{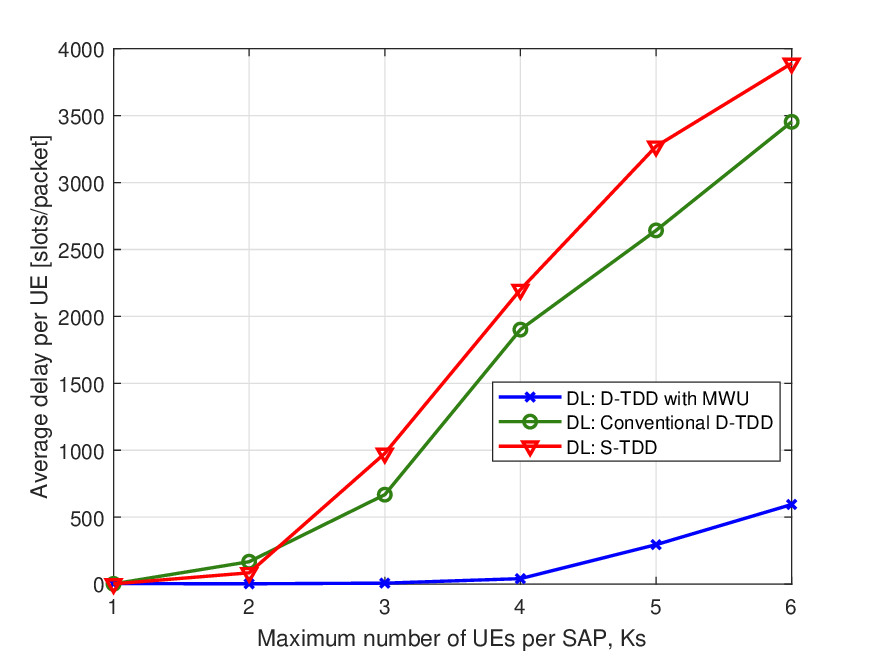}}

  \subfigure[UL transmissions] 
    {\label{fig:delay-UL}
    \includegraphics[width=0.9\columnwidth]{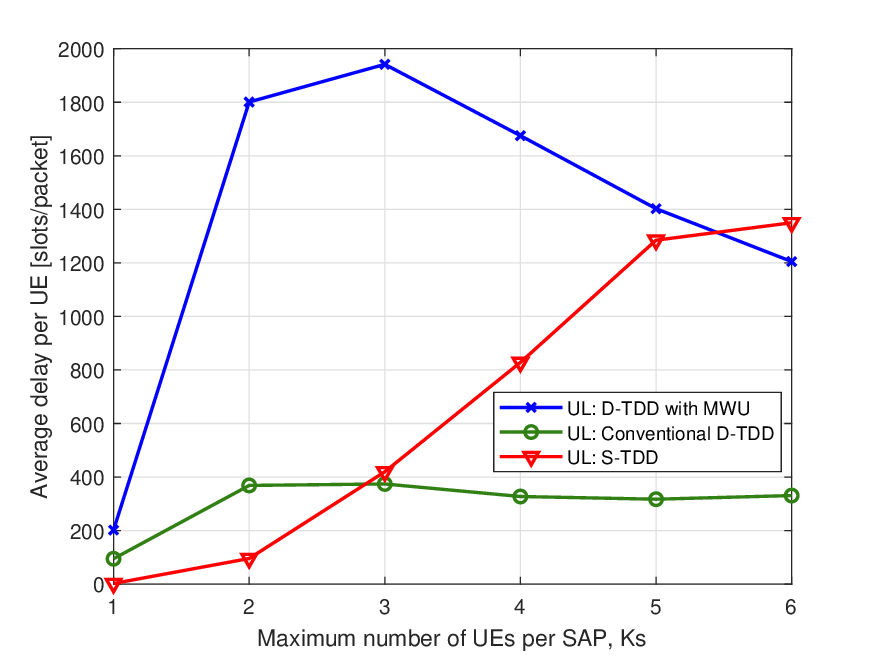}}

  \caption{Mean time delay per UE versus number of UEs per SAP: (a) transmissions in downlink and (b) transmissions in uplink.}

  \label{fig:delay_vs_ue_number}
\end{figure}

\begin{figure}[t!]
  \centering{}

  \subfigure[DL transmissions]
    {\label{fig:Q-DL}
    \includegraphics[width=0.9\columnwidth]{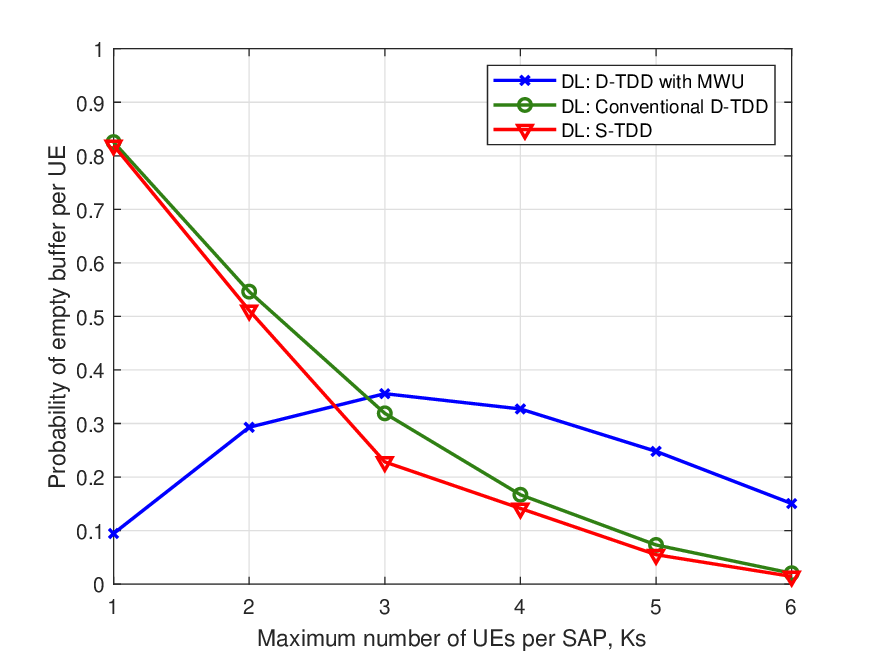}}

  \subfigure[UL transmissions] 
    {\label{fig:Q-UL}
    \includegraphics[width=0.9\columnwidth]{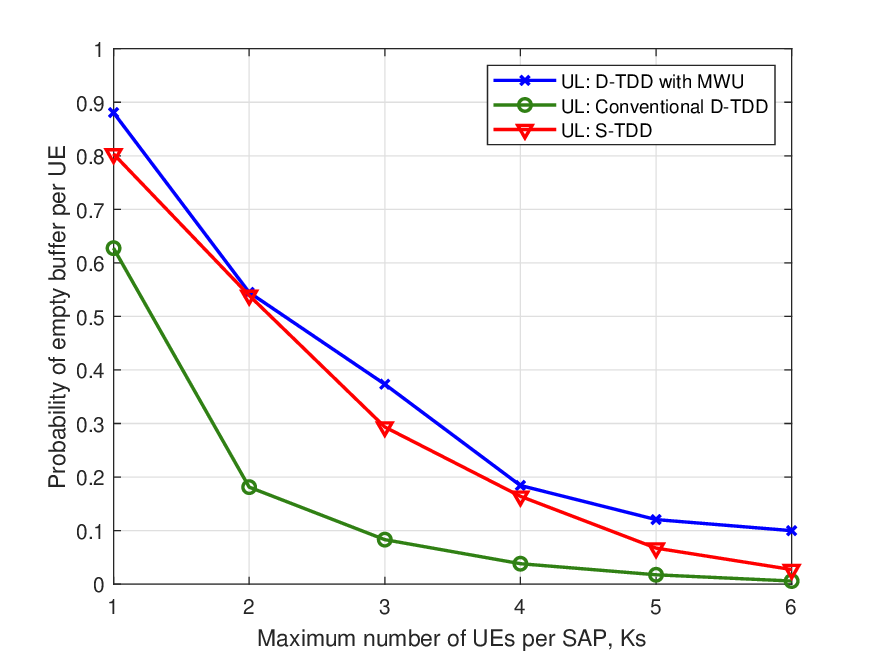}}

  \caption{Pr(Q = 0) at each time slot per UE versus the number of UEs per SAP: (a) transmissions in downlink and (b) transmissions in the uplink.}

  \label{fig:empQ_vs_ue_number}
  
\end{figure}

Fig.~\ref{fig:PcktThrpt_vs_theta} depicts the mean packet throughput per UE as a function of the SIR decoding threshold, under various TDD operation modes.
Fig.~\ref{fig:PcktThrpt_vs_theta}(a) demonstrates that although the conventional D-TDD (under fixed configuration) outperforms S-TDD in the DL transmission, integrating MWU into D-TDD brings about more than two-fold improvement in terms of the mean packet throughput. 
Such a gain is primarily attributed to the capability of the MWU algorithm to adaptively update the time portion allocated to UL and DL based on the feedback of SIR information.
A more striking observation comes from Fig.~\ref{fig:PcktThrpt_vs_theta}(b), which compares the uplink performance among the three TDD schemes. 
This figure shows that ($i$) S-TDD achieves a higher mean packet throughput than the conventional D-TDD in UL, which is in line with the conclusions in \cite{YanGerZho:17} because UL UEs under D-TDD normally encounter more severe interference conditions due to the concurrent UL/DL transmissions; nonetheless, ($ii$) adopting MWU to D-TDD results in an even better UL packet throughput than that under S-TDD. 
This experiment indicates that by adequately aligning the UL/DL transmissions in accordance with the SIR, D-TDD can outperform S-TDD in both transmission directions. 

Fig.~\ref{fig:PcktThrpt_vs_UE_per_cell} investigates the performance of the proposed TDD scheme with an increase in the traffic load (which is reflected by increasing the number of UEs per SAP, $K_{ \mathrm{s} }$).
Fig.~\ref{fig:PcktThrpt_vs_UE_per_cell}(a) shows that regardless of the employed channel access policy, the mean packet throughput decreases drastically with the increase of $K_{ \mathrm{s} }$.
This is because increasing the number of participating UEs in the network not only leads to more incoming packets per cell but, more critically, exacerbates the interference condition. 
Consequently, the UEs' (and also SAPs') activated periods are prolonged, giving rise to deteriorated SIR and extended queue length at every transmitter.
Nevertheless, D-TDD with MWU presents a consistent (as well as remarkable) improvement compared to the S-TDD and conventional D-TDD protocols in DL packet throughput, demonstrating its effectiveness in boosting network performance. 
Fig.~\ref{fig:PcktThrpt_vs_UE_per_cell}(b) demonstrates the gain of D-TDD with MWU in the uplink, showing that applying MWU algorithms in D-TDD achieves an enhancement over S-TDD and conventional D-TDD as well.
To this end, we confirm that ($i$) as the value of $K_{\mathrm{s}}$ increases, the probability for each UE to be selected by its associated SAP at each time slot decreases, leading to a worse throughput performance; ($ii$) the proposed D-TDD with MWU is able to maintain good performance under the influence of $K_{\mathrm{s}}$.

Fig.~\ref{fig:delay_vs_ue_number} plots the average transmission delay per UE as a function of UE number per SAP under different schemes. 
Fig.~\ref{fig:delay_vs_ue_number}(a) shows that for DL transmissions, integrating MWU into D-TDD substantially reduces transmission delay, demonstrating its effectiveness in boosting network performance.
Fig.~\ref{fig:delay_vs_ue_number}(b) shows the instability of D-TDD with MWU compared with the other two schemes in the UL. Nevertheless, the adaptive update of the time portion allocated to UL and DL benefits the adaptation of the traffic load, resulting in a noticeable decline in mean transmission delay.

Fig.~\ref{fig:empQ_vs_ue_number} explores the effects of the proposed algorithm on the queueing dynamics.
Specifically, this figure displays the probability of having empty UL and DL buffers at a typical UE, i.e., $Pr(Q_{x_i}(t)=0)$, at each time slot.
Note that the higher this probability, the more frequently the UE stays in the idle stage, indicating its data packets have been depleted in time. 
Fig.~~\ref{fig:empQ_vs_ue_number}(a) shows that as the traffic load increases, while the $Pr(Q=0)$ under the conventional D-TDD and S-TDD suffer a drastic decrease, the $Pr(Q=0)$ under D-TDD with MWU increases and then decreases. 
According to the feedback of SIR information, the DL transmission probability $\eta_{x_i}(t)$ under D-TDD with MWU increases as the number of UEs per SAP increases, mitigating queues in the buffer.
Fig.~~\ref{fig:empQ_vs_ue_number}(b) demonstrates that D-TDD with MWU enjoys more time in the empty queue, leading to the enhancement over S-TDD and conventional D-TDD.


\section{Conclusion}
In this paper, we devised a decentralized algorithm to update the UL/DL transmission probabilities for D-TDD in small cell networks. 
The proposed method collects the SIR information at each node, which is subsequently input to an MWU-based algorithm to update the time portion allocated to UL and DL at each time slot, aligning the packet transmissions toward the high-quality link directions.
Our simulation results confirm that integrating MWU into D-TDD significantly improves UL and DL mean packet throughput, offering a promising approach for D-TDD transmission protocol designs.

\bibliographystyle{IEEEtran}
\bibliography{bib/StringDefinitions,bib/IEEEabrv,bib/howard_trff_schedule}

\end{document}